\documentclass[showpacs,10pt,twocolumn,prb]{revtex4-1}
\usepackage{amsmath}
\usepackage{amssymb}
\usepackage{graphics}
\usepackage{epsfig}
\usepackage{CJK}
\usepackage{color}

\begin{document}


\title{Physical properties of KMgBi single crystals}
\author{Xiao Zhang$^{1,\dag}$, Shanshan Sun$^{2,\dag}$, and Hechang Lei$^{2,*}$}
\affiliation{$^{1}$State Key Laboratory of Information Photonics and Optical Communications $\&$ School of Science, Beijing University of Posts and Telecommunications, Beijing 100876, China
\\$^{2}$Department of Physics and Beijing Key Laboratory of Opto-electronic Functional Materials $\&$ Micro-nano Devices, Renmin University of China, Beijing 100872, China}

\date{\today}

\begin{abstract}
KMgBi single crystals are grown by using the Bi flux successfully. KMgBi shows semiconducting behavior with a metal-semiconductor transition at high temperature region and a resistivity plateau at low temperature region, suggesting KMgBi could be a topological insulator with a very small band gap. Moreover, KMgBi exhibits multiband feature with strong temperature dependence of carrier concentrations and mobilities.
\end{abstract}


\maketitle


\section{Introduction}

Topological materials have attracted tremendous attentions in the last decade, not only because of their fundamental importance but also because of their potential applications in future technology. After a large number of theoretical and experimental studies, various topological materials have been predicted and confirmed, such as topological insulators (TIs) and superconductors (TSCs),\cite{Hasan1,QiXL1} Dirac semimetals (DSMs),\cite{WangZ1,WangZ2,LiuZK1,LiuZK2} and Weyl semimetals (WSMs).\cite{Nielsen,WanX,XuG,WengH,HuangSM,XuSY,LvBQ,YangLX, HuangX} Moreover, Weyl fermions in the condensed matters can be further classified into two types.\cite{Soluyanov} For type-I WSMs, there is a topologically protected linear crossing of two bands (Weyl point) at the Fermi energy level ($E_{F}$), resulting in a point-like Fermi surface. In contrast, for type-II WSMs, when the Lorentz invariance is violated, the conical spectrum can be strongly tilted and the Fermi surface is no longer point-like. Instead, it consists of electron and hole pockets with finite density of states at $E_{F}$. In recent years, the type-II WSMs have been theoretically predicted and experimentally verified in several materials, like (W, Mo)Te$_{2}$.\cite{Soluyanov,SunY,Kourtis,ChangTR,WangZJ,HuangL,Deng,XuN,LiangA,JiangJ,WangC,WuY,Bruno} Although the type-I DSMs, such as Na$_{3}$Bi and Cd$_{3}$As$_{2}$, have been discovered,\cite{WangZ1,WangZ2,LiuZK1,LiuZK2} the type-II DSMs which are the spin-degenerate counterparts of type-II WSMs are still rare. Very recently, theoretical study predicts that PtSe$_{2}$-type materials are type-II DSMs and it is confirmed by angle-resolved photoemission spectroscopy (ARPES) measurements.\cite{HuangH,YanM}

On the other hand, the PbClF-structure type is an important structural prototype, which has been found in hundreds of materials. They exhibit various novel physical properties, such as superconductivity in Li/NaFeAs and ferromagnetic semiconducting behavior in Li(Zn,Mn)As etc.\cite{Tapp,WangXC,Parker,DengZ} Importantly, recent experimental and theoretical studies further indicate that materials with PbClF-structure could also host various topological states. For example, the monolayer of ZrSiO is predicted to be a two-dimensional (2D) TI and many of isostructural compounds with formula of WHM (W = Zr, Hf, or La, H = Si, Ge, Sn, or Sb, and M = O, S, Se, and Te) possess a similar electronic structure.\cite{XuDN} The ARPES measurements reveal that the topmost unit cell on the (001) surface of ZrSnTe single crystals could be a 2D TI with a curved $E_{F}$,\cite{LouR} thus confirming the theoretical prediction. Moreover, if the spin-orbit coupling (SOC) can be neglected, they will become three-dimensional (3D) node-line DSMs, which has been observed in ZrSiS.\cite{Schoop}

Very recently, KMgBi with PbClF-structure is theoretically predicted to be a 3D DSM and the Dirac fermions are protected by the $C_{4v}$ point group symmetry.\cite{LeC} Furthermore, when compared to Na$_{3}$Bi and Cd$_{3}$As$_{2}$, the dispersion of the Dirac fermions in KMgBi is highly anisotropic because of weak dispersion of the $p_{x}$ and $p_{y}$ orbitals of Bi along the $z$-direction. It suggests that KMgBi is located at the edge of type-I and type-II DSM phases. When doping Rb or Cs into K site, K$_{1-x}$R$_{x}$MgBi(R = Rb, Cs) can be tuned between these two DSM phases.\cite{LeC} Motivated by the theoretical study, in this work, we study the physical properties of KMgBi single crystals in detail. Experimental results indicate that KMgBi exhibits a narrow-band semiconducting behavior with multiband feature, different from the theoretically predicted 3D DSM behavior. Moreover, there is a resistivity plateau appearing at low temperature region, implying that there may be a nontrivial topological surface state in KMgBi. Interestingly, a non-bulk superconducting transition emerges at $T_{c}\sim$ 2.8 K. It could originate either from extrinsic superconducting second phase or from intrinsic filamentary superconductivity of KMgBi.

\section{Experimental}

KMgBi single crystals were grown by using the Bi flux with the molar ratio of K : Mg : Bi = 1 : 1 : 18. K chunk (99.5 $\%$), Mg chip (99.9 $\%$) and Bi shot (99.99 $\%$) were mixed and put into an alumina crucible, covered with quartz wool and then sealed into the quartz tube with partial pressure of argon. The sealed quartz ampoule was heated to and soaked at 600 K for 4 h, then cooled down to 300 K with 3 K/h. At this temperature, the ampoule was taken out from the furnace and decanted with a centrifuge to separate KMgBi crystals from Bi flux. KMgBi single crystals with typical size 3$\times$4$\times$0.3 mm$^{3}$ can be obtained. Because the raw materials and KMgBi are highly air-sensitive, all manipulations were carried out in an argon-filled glovebox with an O$_{2}$ and H$_{2}$O content below 0.1 ppm. X-ray diffraction (XRD) patterns of powdered small crystals and a single crystal were collected using a Bruker D8 X-ray Diffractometer with Cu $K_{\alpha}$ radiation ($\lambda=$ 0.15418 nm) at room temperature. Rietveld refinements of the XRD patterns were performed using the code TOPAS4.\cite{TOPAS} Electrical transport and heat capacity measurements were carried out in a Quantum Design PPMS. The longitudinal and Hall electrical resistivity were measured using a four-probe method on single crystals cutting into rectangular shape. The current flows in the $ab$ plane of samples. The Hall resistivity was obtained from the difference of the transverse resistivity measured at the positive and negative fields in order to remove the longitudinal resistivity contribution due to voltage probe misalignment, i.e., $\rho_{xy}(H)=[\rho(+H)-\rho(-H)]/2$. Magnetization measurements were carried out by using a Quantum Design MPMS3.

\section{Results and discussion}

\begin{figure}[tbp]
\centerline{\includegraphics[scale=0.38]{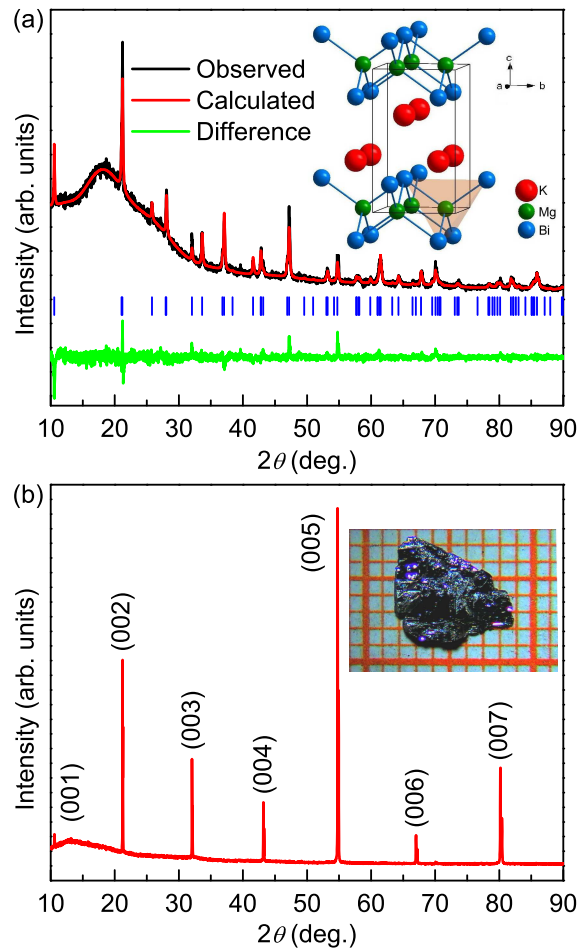}} \vspace*{-0.3cm}
\caption{(a) Powder XRD pattern of KMgBi. Inset: Crystal structure of KMgBi. The small green, medium blue and big red balls represent Mg, Bi and K atoms, respectively. (b) XRD pattern of a KMgBi single crystal. Inset: photo of a typical KMgBi single crystal. The length of one grid in the photo is 1 mm.}
\end{figure}

The crystal structure of KMgBi is composed of the Mg-Bi and K layers stacking along the $c$-axis direction alternatively (inset of Fig. 1(a)).\cite{Vogel} In Mg-Bi layer, Each Mg atom is coordinated with four Bi atoms and the tetrahedra of MgBi$_{4}$ are connected each other by edge-sharing. The crystal structure of KMgBi with the space group $P4/nmm$ is similar to the 111 family of iron-based superconductors,\cite{Tapp,WangXC} where K, Mg and Bi atoms are replaced by Li, Fe and As atoms, respectively. The main panel of Fig. 1(a) shows the powder XRD pattern and refinement of crushed KMgBi crystals. All reflections can be well indexed using the $P4/nmm$ space group. The determined lattice parameters are $a=$ 4.8808(4) \r{A}\ and $c=$ 8.3765(3) \r{A}, consistent with previous results.\cite{Vogel} The XRD pattern of a single crystal (Fig. 1(b)) reveals that the crystal surface is normal to the $c$-axis with the plate-shaped surface parallel to the $ab$-plane. The plate-like crystals with square shape (inset of Fig. 1(b)) is consistent with the layered structure and the tetrahedron symmetry of KMgBi. Moreover, the crystals are rather soft and very easy to cleave along the $ab$-plane.

\begin{figure}[tbp]
\centerline{\includegraphics[scale=0.42]{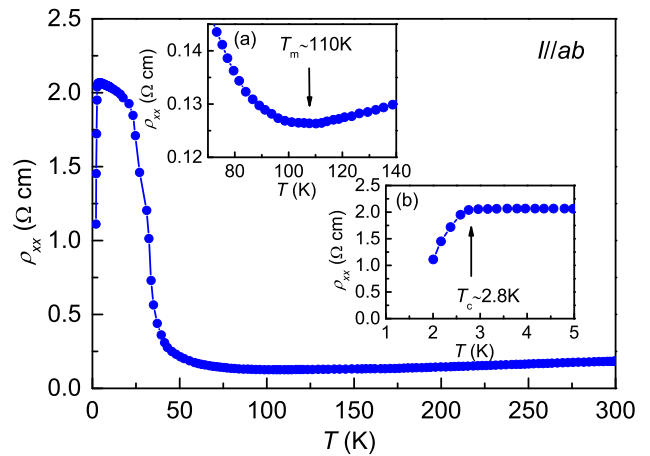}} \vspace*{-0.3cm}
\caption{Temperature dependence of the in-plane resistivity $\rho_{xx}(T)$ of KMgBi single crystal at zero field. Inset: enlarged part of $\rho_{xx}(T)$ (a) between 70 and 140 K and (b) below 5 K.}
\end{figure}

Temperature dependence of zero-field electrical resistivity $\rho_{xx}(T)$ in the $ab$-plane for KMgBi single crystal is shown in Fig. 2. The $\rho_{xx}(T)$ decreases with lowering temperature gradually and then exhibits semiconducting behavior when $T<T_{m}$ ($\sim$ 110 K) (inset (a) of Fig. 2). Moreover, the absolute value of $\rho_{xx}(T)$ is relative small. Thus, it suggests that KMgBi should be a narrow-band semiconductor rather than a semimetal. Assuming the semiconducting behavior is intrinsic and using the thermal activation model $\rho_{xx}(T)=\rho_{0}$exp($E_{g}/2k_{B}T$), the fit of $\rho_{xx}(T)$ between 40 and 100 K gives the band gap $E_{g}\sim$ 11.2(3) meV (130(4) K), which is much smaller than the calculated $E_{g}$ without SOC (362.8 meV).\cite{LeC} On the other hand, the metal-semiconductor transition at high temperature region can also be explained by this small band gap: once the temperature is comparable to the $E_{g}$, it will become degenerate semiconductor behaving like a bad metal. This phenomenon has been observed in some of doped semiconductors, such as Ga doped ZnO films and indium tin oxide films etc.\cite{Bhosle,LiY}

Surprisedly, when further decreasing temperature below about 20 K, the slope of $\rho_{xx}(T)$ curve becomes smaller and the resistivity plateau appears. It implies that another conducting channel becomes dominant when the bulk resistance becomes relatively high. This phenomenon is very similar to those observed in TIs, like Bi$_{2}$Te$_{2}$Se and SmB$_{6}$,\cite{RenZ,KimDJ} and could be explained by the existence of metallic surface state which is topologically protected. This surface state has been predicted in theory.\cite{LeC} Moreover, theoretical calculation indicates that when compressing the lattice along the $a$ axis with 2\% change, the Dirac nodes in KMgBi will be gapped out and KMgBi will change from a DSM to a three-dimensional (3D) strong TI.\cite{LeC} It seems more consistent with the present $\rho_{xx}(T)$ result. This surface state needs to be confirmed further by other techniques, such as ARPES measurement. More interestingly, there is a drop in the $\rho_{xx}(T)$ curve when $T<$ 2.8 K (inset (b) of Fig. 2) and this transition shifts to lower temperature under magnetic field (not shown here). It indicates that this drop results from a superconducting transition. Due to the broad transition width and the non-zero $\rho_{xx}(T)$ at 2 K, this superconductivity should be non-bulk, which is confirmed by heat capacity and magnetization measurements shown below.

\begin{figure}[tbp]
\centerline{\includegraphics[scale=0.35]{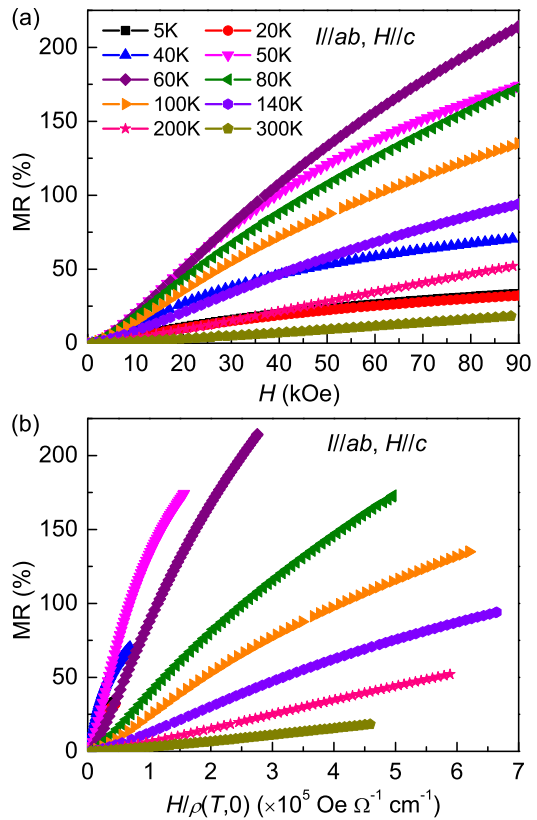}} \vspace*{-0.3cm}
\caption{(a) Magnetoresistance (MR) of $\rho_{xx}(T,H)$ and (b) MR vs. $H/\rho_{xx}(T,0)$ at various temperatures for $H\parallel c$.}
\end{figure}

The magnetoresistance (MR) (MR $=(\rho_{xx}(T,H)-\rho_{xx}(T,0))/\rho_{xx}(T,0)=\Delta\rho_{xx}/\rho_{xx}(T,0)$) at various temperatures are shown in Fig. 3(a). The MR at 5 K is about 33 \% under $H=$ 90 kOe, which is much smaller than topological or compensated SMs, such as Cd$_{3}$As$_{2}$, TaAs, WTe$_{2}$, LaBi etc.\cite{LiangT,HuangX,Ali,SunSS2} Moreover, the MR at $H=$ 90 kOe does not change with temperature monotonically. With increasing temperature, the MR increases initially and then starts to decrease. The maximum of MR appears at $T\sim$ 60 K. The small MR at low temperatures can be partially related to the rather large $\rho_{xx}(T,0)$. On the other hand, if there is a single type of carrier and the scattering time $\tau$ is same at all points on the Fermi surface, the MR will follow the Kohler's rule: the MR measured at various temperatures can be scaled into a single curve, MR $=F(H/\rho_{xx}(T,0))$.\cite{Pippard,Ziman} Clearly, the scaled MR curves do not fall into one curve for KMgBi (Fig. 3(b)), indicating that the Kohler's rule is violated. It strongly suggests that there are more than one type of carrier and the carrier concentrations and/or the mobilities of electron and hole are strongly temperature dependent.\cite{McKenzie,ZhanY}

\begin{figure}[tbp]
\centerline{\includegraphics[scale=0.42]{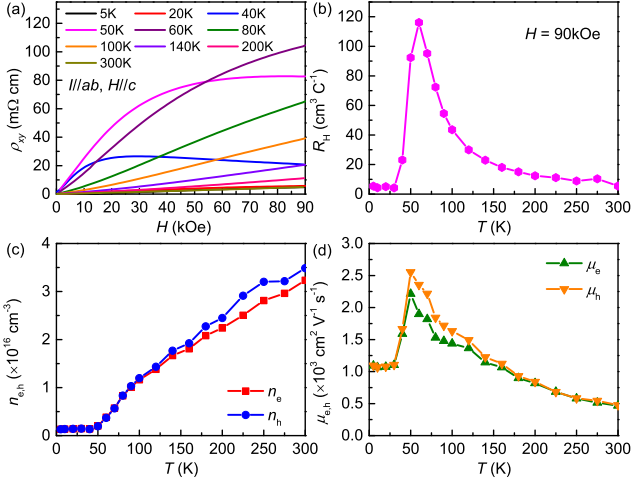}} \vspace*{-0.3cm}
\caption{(a) Field dependence of Hall resistivity $\rho_{xy}(H)$ up to $H$ = 90 kOe for $H\Vert c$ at various temperatures. (b) Temperature dependence of Hall coefficient $R_{H}(T)$ at $H$ = 90 kOe. (c) Temperature dependence of fitted electron- and hole-type carrier concentrations $n_{e,h}(T)$. (d) Fitted mobilities of electron and hole $\mu_{e,h}(T)$ as a function of temperature.}
\end{figure}

Fig. 4(a) shows the field dependence of Hall resistivity $\rho_{xy}(H)$ at various temperatures. The $\rho_{xy}(H)$ curves are almost linear with positive slope at high temperatures, and start to bend strongly when $T<$ 60 K. With further decreasing temperature ($T\leq$ 30 K), the $\rho_{xy}(H)$ curves become nearly linear again and the slopes are small. The corresponding Hall coefficient $R_{H}(T)$ ($=\rho_{xy}(T,H)/H$) at $H$ = 90 kOe increases with decreasing temperature at first and then decreases quickly when $T<$ 60 K. Finally, at $T\leq$ 30 K, the $R_{H}(T)$ becomes almost temperature independent. Interestingly, the temperature region where the $\rho_{xy}(H)$ exhibits remarkably nonlinear behavior accompanying with the sharp drop of $R_{H}(T)$ is nearly same as that where the $\rho_{xx}(T)$ shows remarkably semiconducting behavior.

According to the two-band model, the $\rho_{xy}(H)$ can be expressed as,\cite{Pippard,Ziman}

\begin{equation}
\rho_{xy}(H) = \frac{\mu_{0}H}{|e|}\frac{(n_{h}\mu_{h}^{2}-n_{e}\mu_{e}^{2})+(n_{h}-n_{e})(\mu_{e}\mu_{h})^{2}(\mu_{0}H)^{2}}{(n_{h}\mu_{h}+n_{e}\mu_{e})^{2}+(n_{h}-n_{e})^{2}(\mu _{e}\mu _{h})^{2}(\mu_{0}H)^{2}}
\end{equation}

where $n_{e,h}$ and $\mu_{e,h}$ are carrier concentrations and mobilities of electron and hole, respectively. Using this model, the fitted $n_{e,h}$ and $\mu_{e,h}$ as a function of temperature are shown in Fig. 4(c) and (d). The $n_{e,h}(T)$ decrease gradually with lowering temperature but the slope is larger when $T\leq$ 100 K (Fig. 4(c)). Then, the $n_{e,h}(T)$ become almost temperature independent at $T<$ 60 K. The ratio of $n_{h}$ to $n_{e}$ at $T>$ 60 K is larger than 1, indicating that the dominant carrier in KMgBi is hole and consistent with the positive slope of $\rho_{xy}(H)$ at high temperature. In contrast, the $n_{h}/n_{e}$ is slightly smaller than 1 between 20 and 60 K, i.e., there are more electrons than holes, leading to the strong downward bending of $\rho_{xy}(H)$ at high field region. The estimated $n_{e,h}$ at 2 K and 300 K are about 1.3$\times$10$^{15}$ and 3.3$\times$10$^{16}$ cm$^{-3}$, respectively. Such low $n_{e,h}$ are comparable with those in typical TIs, such as Bi$_{2}$Te$_{2}$Se and Sb doped Bi$_{2}$Se$_{3}$,\cite{RenZ,Analytis} but much smaller than those in SMs, such as Cd$_{3}$As$_{2}$ and WTe$_{2}$.\cite{LiangT,LuoY} Moreover, the strong enhancement of $n_{e,h}(T)$ with temperature at $T>$ 50 K also confirms the semiconducting feature of KMgBi. On the other hand, the $\mu_{e,h}(T)$ exhibit similar temperature dependence to the $R_{H}(T)$. They increase monotonically with decreasing temperature from 300 K to 60 K and this is often observed when the lattice scattering is dominant. In contrast, the $\mu_{e,h}(T)$ decrease quickly when $T<$ 60 K and then become nearly temperature independent below 30 K. The decreases of $\mu_{e,h}(T)$ should be the main cause of the remarkable increase of $\rho_{xx}(T)$ and it could be due to the impurity scattering, which enhances with decreasing temperature and dominates over the lattice scattering at low temperature region.\cite{Pierret} Moreover, strong temperature dependence of $n_{e,h}(T)$ and $\mu_{e,h}(T)$ should be the reasons that cause the violation of Kohler's rule in KMgBi (Fig. 3(a)).

\begin{figure}[tbp]
\centerline{\includegraphics[scale=0.35]{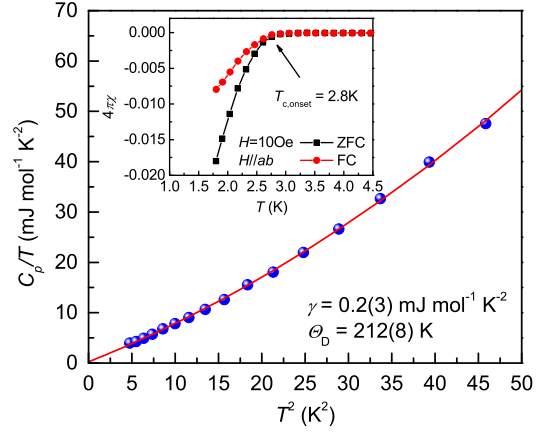}} \vspace*{-0.3cm}
\caption{The relationship between $C_{p}/T$ and $T^{2}$ for KMgBi single crystal at low temperature region. The red solid curve represents the fitting result using the formula $C_{p}/T=\gamma+\beta T^{2}+\delta T^{4}$. Inset: temperature dependence of magnetic susceptibility $\chi(T)$ at $H=$ 10 Oe with ZFC and FC modes for $H\parallel ab$.}
\end{figure}

Fig. 5 shows the temperature dependence of heat capacity for KMgBi single crystal. The relation between $C_{p}(T)/T$ and $T^{2}$ is not linear even at very low temperature region (2.2 K $<T<$ 6.8 K). The fit using the formula $C_{p}/T=\gamma+\beta T^{2}+\delta T^{4}$ is rather good and it gives $\gamma$ = 0.2(3) mJ mol$^{-1}$ K$^{-2}$, $\beta$ = 0.68(3) mJ mol$^{-1}$ K$^{-4}$, and $\delta$ = 0.0079(6) mJ mol$^{-1}$ K$^{-6}$. The electronic specific heat coefficient $\gamma$ is almost zero, consistent with the extremely low carrier concentrations of KMgBi from Hall measurement. The Debye temperature $\Theta _{D}$ is estimated to be 212(8) K using the formula $\Theta _{D}$ = $(12\pi^{4}NR/5\beta )^{1/3}$, where $N$ is the atomic number in the chemical formula ($N$ = 3) and $R$ is the gas constant ($R$ = 8.314 J mol$^{-1} $ K$^{-1}$). The absence of the heat capacity jump at $T\sim$ 2.8 K indicates that the superconductivity appearing in KMgBi is not bulk. This non-bulk superconductivity is confirmed by the magnetization measurement (inset of Fig. 5). Although there is a diamagnetic transition in the magnetic susceptibility $\chi(T)$ curves at $T_{c,onset}\sim$ 2.8 K when $H=$ 10 Oe with zero-field-cooling (ZFC) and field-cooling (FC) modes and the transition temperature is close to that corresponding to the drop in the $\rho_{xx}(T)$ curve, the superconducting volume fractions are very small ($\sim$ 1 and 2 \% at 1.8 K for ZFC and FC $\chi(T)$ curves). Because the $T_{c}$ is close to that of KBi$_{2}$ ($T_{c}=$ 3.57 K),\cite{SunSS} this superconducting phase could originate from the tiny of KBi$_{2}$. But it has to be mentioned that we measured at least ten samples that are carefully cleaved, cut and scratched, and all of samples exhibit superconducting transition with similar $T_{c}$. It implies that there is either a tiny KBi$_{2}$ embedded or an intrinsic filamentary superconductivity existing in KMgBi single crystals. If it is intrinsic, the measurements using the surface-sensitive spectroscopy techniques, such as ARPES or scanning tunneling microscope measurements, would further help to distinguish whether the superconductivity is related to the possible non-trivial surface state.

\section{Conclusion}

In summary, we grow the KMgBi single crystals by using the Bi flux successfully. It has a layered structure with space group $P4/nmm$ (PbClF-structure). KMgBi exhibits the metal-semiconductor transition at $T_{m}\sim$ 110 K with a resistivity plateau below 30 K. It strongly suggests that KMgBi should be a TI with small band gap, which is different from the predicted DSM state in theory. The MR and Hall measurements indicate that there are two kinds of carriers in KMgBi with extremely low concentrations. Moreover, the MR does not follow the Kohler's rule because the carrier concentrations and mobilities of electron and hole significantly depend on temperature. Further analysis implies that the remarkable upturn behavior of $\rho_{xx}(T)$ can be ascribed to the decrease of mobilities. This decrease possibly results from the crossover from lattice scattering to impurity scattering mechanism at low temperature region.

\section{Acknowledgments}

This work was supported by the Ministry of Science and Technology of China (2016YFA0300504), the Fundamental Research Funds for the Central Universities, and the Research Funds of Renmin University of China (RUC) (15XNLF06, 15XNLQ07), and the National Natural Science Foundation of China (Grant No. 11574394).

$\dag$ These authors contributed equally to this work.

$\ast$ hlei@ruc.edu.cn

\end{document}